\documentclass[conference,a4paper,table]{APSIPA2025}
\usepackage{amsmath}
\usepackage{graphicx}
\usepackage{multirow}
\usepackage{threeparttable}
\usepackage[backend=biber,style=ieee,]{biblatex}
\usepackage{balance}
\usepackage{amsmath}
\usepackage{amssymb}
\usepackage{amsfonts}

\usepackage{booktabs}
\usepackage{array}

\usepackage{microtype}
\usepackage{geometry}
\usepackage{fancyhdr}

\fancypagestyle{firststyle}{
  \fancyhf{}
}

\begin{document}

\title{Ensemble Confidence Calibration for Sound Event Detection in Open-environment}

\author{
\authorblockN{
Yuanjian Chen\authorrefmark{1} and
Han Yin\authorrefmark{2}
}

\authorblockA{
\authorrefmark{1}
School of Computer Science and Technology, Harbin University of Science and Technology, Harbin, China \\
E-mail: 2010400002@stu.hrbust.edu.cn}

\authorblockA{
\authorrefmark{2}
School of Electrical Engineering, KAIST, Daejeon, Republic of Korea\\
E-mail: hanyin@kaist.ac.kr}
}

\maketitle
\thispagestyle{firststyle}
\pagestyle{fancy}

\begin{abstract}
Sound event detection (SED) has made strong progress in controlled environments with clear event categories. However, real-world applications often take place in open environments. In such cases, current methods often produce predictions with too much confidence and lack proper ways to measure uncertainty. This limits their ability to adapt and perform well in new situations. To solve this problem, we are the first to use ensemble methods in SED to improve robustness against out-of-domain (OOD) inputs. We propose a confidence calibration method called Energy-based Open-World Softmax (EOW-Softmax), which helps the system better handle uncertainty in unknown scenes. We further apply EOW-Softmax to sound occurrence and overlap detection (SOD) by adjusting the prediction. In this way, the model becomes more adaptable while keeping its ability to detect overlapping events. Experiments show that our method improves performance in open environments. It reduces overconfidence and increases the ability to handle OOD situations.
\end{abstract}

\section{Introduction}
Sound event detection (SED), which involves recognizing event categories and identifying their corresponding timestamps \cite{mesaros2021sound,zhang2024unified}, is an important method for analyzing and interpreting acoustic information within audio clips. SED has been widely applied across various domains, including industrial monitoring \cite{wu2023unsupervised}, healthcare \cite{qiu2024heart}, and smart city development \cite{tan2021extracting}.

Recently, with the flourishing development of the Detection and Classification of Acoustic Scenes and Events (DCASE) challenge series \cite{mesaros2025decade}, a substantial number of excellent research achievements \cite{li2023ast,bhosale2024diffsed,shao2024fine,10887631,yue2025full,guan2024,audiosep-dp} have continuously emerged. Among these advances, there are ensemble model approaches \cite{guo2019multi,wang2021model,afendi2022sound,dinkel2022large,mukhamadiyev2025ensemble}  that integrate the sound event posterior probabilities from individual models to obtain final prediction results. 
The success of current methods mostly depends on the assumption that sound events come from a small, known, and clearly defined set. However, as the need for real-world applications continues to grow and as tasks become more complex, this assumption no longer holds~\cite{zhou2022open,wilddesed}. In open-environment, SED system's performance is significantly affected by factors such as ambiguous event definitions, long-tail data distributions \cite{zhang2024unified,xiao2024mixstyle}, and complex environmental noise, making the learning process more dynamic and uncertain. 

To address open-environment challenges, researchers have conducted several explorations. Wei et al. \cite{wei2020crnn} employed data manipulation-based methods, handling differences between different acoustic environments through domain adaptation techniques. Xiao et al. \cite{10887631} specifically addressed incremental learning challenges through an unsupervised class incremental learning framework. Despite the progress these methods have achieved in their respective targeted challenges, significant limitations remain. First, existing methods lack effective mechanisms for handling the coupling relationships among multiple sound events in open-environment. Second, when confronted with out-of-domain (OOD) challenges, models often generate overconfident predictions and lack effective confidence calibration mechanisms, severely constraining their adaptability and generalization performance in OOD scenes.

To overcome these issues, it is necessary to look at ideas that have worked in related audio tasks. Recent research~\cite{kwok2025robust} has demonstrated that constructing speech anti-spoofing model ensembles and employing linear fusion algorithms to combine classification scores from multiple candidate models can effectively enhance the robustness of classification models in OOD scenes. Following this line, in this work, we introduce the ensemble methods into open-environment SED to address the OOD challenges. However, confidence calibration, as a key technique for reflecting the reliability of model predictions \cite{guo2017calibration}, faces severe challenges in OOD scenarios: OOD inputs often lead models to produce overconfident predictions, thereby compromising the effectiveness of ensemble fusion. 

To solve the overconfidence issue in OOD scenes, we propose a confidence calibration method based on Energy-based Open-World Softmax (EOW-Softmax)~\cite{wang2021energy}, which models uncertainty in open-world conditions. This method is applied before final decision-making to adjust the magnitude of logit scores, allowing better prediction confidence for each class. Considering the polyphonic nature of SED, we apply this technique to sound occurrence and overlap detection~\cite{guan2024}, a statistical sub-task of SED that captures overlapping sound patterns. Our work contributes to the field in three ways: (1) we are the first to introduce ensemble methods into SED for open environments, improving robustness and generalization in OOD scenes using interpolation-based fusion; (2) we address the common overconfidence problem in OOD predictions by using EOW-Softmax to calibrate model confidence based on open-world uncertainty; and (3) we design a task-specific strategy for applying EOW-Softmax to polyphonic detection by adjusting logit magnitudes, ensuring that ensemble models remain adaptable without losing their ability to detect overlapping events. The code will be publicly available soon.

\section{Methods}
\subsection{Out-of-domain challenge for SED}
As we mentioned, the SED systems perform exceptionally well in controlled settings but experience considerable difficulties in open environments, such as those encountered in intelligent security surveillance, where acoustic scenes vary dramatically between indoor and outdoor environments, and event definitions are often ambiguous. To model the evolution of the dataset across different domains, we introduce $\mathbb{D}^{1:s}_{train}=\left \{ \mathbb{X}^{1:s}_{train}, \mathbb{Y}^{1:s}_{train} \right \}$ for data up to domain $s$ during training stage, where $\mathbb{X}^{1:s}_{train}$ denotes the set of audio samples and $\mathbb{Y}^{1:s}_{train}$ represents the corresponding frame-level annotations, $\mathbb{D}^{s+1}_{test}=\left \{ \mathbb{X}^{s+1}_{test}, \mathbb{Y}^{s+1}_{test} \right \}$ for the subsequent domain $s+1$ during testing stage. The OOD challenge involves the model's ability to generalize effectively across significantly different domains. While leveraging foundational models can partially address OOD challenges, it may lead to overfitting or performance degradation in new domains \cite{liang2024object}. As shown in Fig.\ref{fig:fig1}, the issue of OOD challenge lies in the distributional gap between training data $\mathbb{X}^{1:s}_{train}$ and testing data $\mathbb{X}^{s+1}_{test}$, represented as $P(\mathbb{X}^{1:s}_{train})\ne P(\mathbb{X}^{s+1}_{test})$. The desired robustness property, $P(\mathbb{Y}^{1:s}_{train}\mid \mathbb{X}^{1:s}_{train})=P(\mathbb{Y}^{s+1}_{test}\mid\mathbb{X}^{s+1}_{test})$, captures the expectation that conditional relationships learned from source scenes should generalize to target scenes. Assuming that our training set can be expressed as $\left \{ (x_{n},y_{n}) \right \}_{n=1}^{N}$, where $x_{n}$ denotes the audio of the $n$-th clips $\mathbb{X}^{1:s}_{train}$, with frame-level label $y_{n}=\left \{y^{1}_{n},...,y^{t}_{n},...,y^{T}_{n}\right \}$ forming a sequence of $C$-dimensional binary vectors $y^{t}_{n}\in \left \{ 0,1 \right \}_{C}$ for $C$ event categories across $T$ temporal frames, $N$ represents the total clips in the training set. The SED model $\mathcal{F}_{\phi }$ aims to optimize its performance by learning the parameter $\phi$. Thus, the optimization objective can be expressed as:

\begin{figure}[t!]
\centering
\includegraphics[width=0.75\columnwidth]{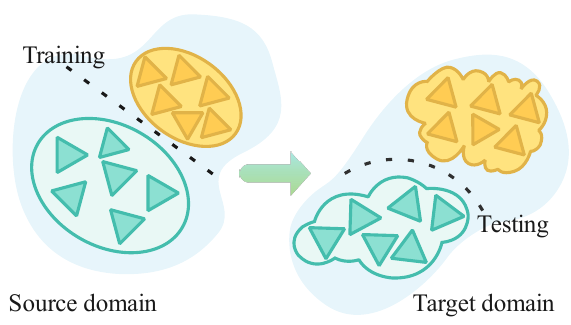}
\vspace{-5pt}
\caption{Overview of OOD challenge for SED. Training in source domain acoustic scenes and testing in target domain acoustic scenes.}
\label{fig:fig1}
\vspace{-10pt}
\end{figure}

\begin{equation}
min\sum_{n=1}^{N}\mathcal{L}_{\text{SED}}(\mathcal{F}_{\phi }(x_{i}),y_{i})+\tau \mathcal{L}_{\text{OPEN}}
\label{eq:openobj}
\end{equation}

\noindent where $\mathcal{L}_{\text{SED}}$ denotes the standard SED objective function, while $\mathcal{L}_{\text{OPEN}}$ represents an auxiliary loss term designed to handle open-environment challenges, with $\tau$ serving as the weighting hyperparameter.

\subsection{Candidate Model and EOW-Softmax}

\begin{figure*}[t]
\centering
\includegraphics[width=0.7\textwidth]{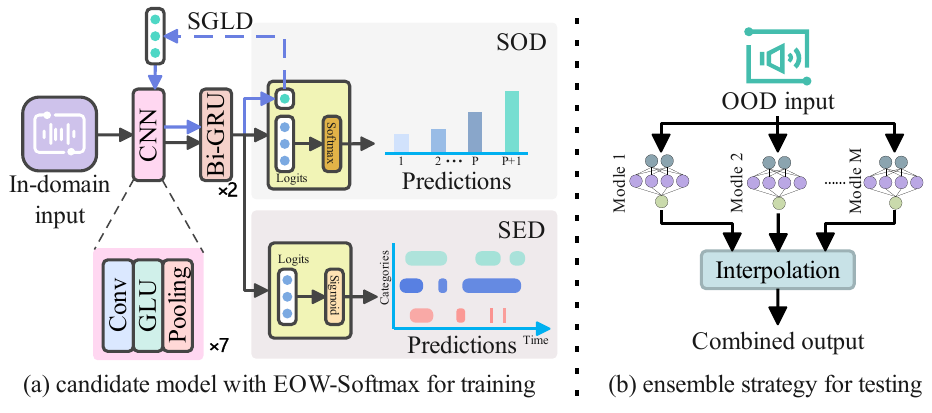}
\vspace{-5pt}
\caption{Pipeline of the proposed approach: (a) dual-branch candidate model for training with SED and SOD tasks, and (b) model ensemble strategy with interpolation for testing.}
\label{fig:fig2}
\vspace{-10pt}
\end{figure*}

We adopt the baseline of DCASE 2023 SED challenge as our candidate model. To better address polyphonic phenomena in open-environment and improve event boundary detection performance, we introduce the sound occurrence and overlap detection (SOD) task \cite{guan2024}. This task focuses on sound activity patterns rather than specific event categories, providing more effective frame-level supervision signals for SED. The overall framework is illustrated in Figure~\ref{fig:fig2}. This framework consists of four components: seven CNN blocks, two Bi-GRU blocks, and two prediction heads for SED and SOD tasks, respectively . The prediction head of SED outputs frame-level event predictions, while the SOD branch outputs frame-level event activity states categorized into three classes: $\left \{ 0,1,2\right\}$, where $0$ indicates no target events, $1$ represents monophonic event, and $2$ denotes polyphonic events.

In the SED branch, we adopt binary cross entropy (BCE) as $\mathcal{L}_{SED}$ in our equation \ref{eq:openobj}. To address OOD challenges in SED and perform confidence calibration for the candidate model, we employ the EOW-Softmax \cite{wang2021energy} strategy to train the model in the SOD branch, thereby improving the output fusion of model ensemble. The EOW-Softmax strategy addresses open-world uncertainty as an additional dimension \cite{wang2021energy}, implementing a $(P+1)$-way classification scheme. Here, the primary $P=3$ dimensions encode the three event activity categories within the SOD branch, whereas the extra fourth dimension quantifies the open-world uncertainty. This uncertainty metric increases when the model exhibits low prediction confidence. Following Figure~\ref{fig:fig2}(a), let $\Upsilon _{\phi }:\mathbb{R}^{D}\to \mathbb{R}^{P+1}$ denotes the linear mapping in the SOD prediction head, yielding four logits expressed as $\Upsilon _{\phi }(x)\left \| i \right \| $, with $i \in \left \{ 1,2,3,4 \right \} $, for the input in-domain acoustic representation $x \in \mathbb{R}^{D} $. Probability distributions are obtained via softmax transformation:

\begin{equation}
\Psi _{\phi }(x)\left \| i \right \|=\frac{\exp \Upsilon _{\phi }(x)\left \| i \right \| }{\sum_{j=1}^{P+1}\exp \Upsilon _{\phi }(x)\left \| j \right \|} 
\label{eq:softnorm}
\end{equation}
where $\Psi _{\phi }$ is the concatenation of the linear layer and a softmax transformation layer. Then, the loss function $\mathcal{L}_{OPEN}(x,\mathfrak{o})$ for Eow-Softmax is defined as:

\begin{equation}
\begin{split}
\mathcal{L}_{\text{OPEN}}(x,\mathfrak{o})=&\underset{\phi}{\min}\mathbb{E}_{p(x)}\left [ -\log{\Psi _{\phi }(x)\left \| \mathfrak{o} \right \|}  \right ] \\
&\hspace{-8.9mm} + \lambda \mathbb{E}_{p_{\hat{\phi}}(x)}\left [ -\log{\Psi _{\phi }(x)\left \| P+1 \right \|}  \right ]
\end{split}
\label{eq:leow}
\end{equation}
where $\lambda \in \mathbb{R}^+$ is a hyperparameter. The initial component represents the maximum log-likelihood (MLL) criterion for the SOD task utilizing ground truth labels $\mathfrak{o}$; the subsequent component constitutes the MLL objective for identifying acoustic features sampled from normal distributions $p_{\hat{\phi}}(x)$, where $\hat{\phi}$ represents the frozen parameters from $p_{\phi}(x)$ at the current iteration. These normal distributions are learned through SGLD-based optimizations \cite{wang2021energy}. As a result, our candidate model should be optimized to generate elevated open-world uncertainty when processing samples from the latter distribution.

\subsection{Model Ensemble Strategy}
To enhance SED performance in open-environment, we adopt confidence-based model ensemble approach. As shown in Figure~\ref{fig:fig2}(b), multiple candidate models' results are combined to enhance the ensemble's prediction. Although confidence calibration strategy is incorporated into the SOD sub-task during training, we focus on optimizing SED task performance during the testing stage. As the confidence calibration from the SOD branch encodes the model's frame-level audio comprehension accuracy, we leverage SOD confidence to guide the SED ensemble process. For the $c$-th event category, the final SED combined prediction $s_{\text{SED}}\left ( x \right ) \left [ c \right ] $ is defined as: 

\begin{equation}
s_{\text{SED}}\left ( x \right ) \left [ c \right ] =\frac{\sum_{m=1}^{M} \mathcal{F}_{\phi_{m}}\left ( x \right )\left [ c \right ] \cdot c_{m}(x) }{\sum_{m=1}^{M} c_{m}(x)}  
\label{eq:ensemble}
\end{equation}
where $\mathcal{F}_{\phi_{m}}\left ( x \right )\left [ c \right ]$ denotes the prediction of the SED branch of the $m$-th candidate model for the $c$-th event category, $c_m(x)$ is the confidence calibration learned by the SOD branch of the $m$-th candidate model, and $M$ is the total number of candidate models in the ensemble. The SOD confidence serves as weights to modulate each candidate model's contribution. Models with higher confidence play a more significant role in final predictions, thereby improving SED performance in open-environment.

\section{Experiments}

\subsection{Dataset and Performance Metric}

\begin{figure}[t!]
\centering
\includegraphics[width=\columnwidth]{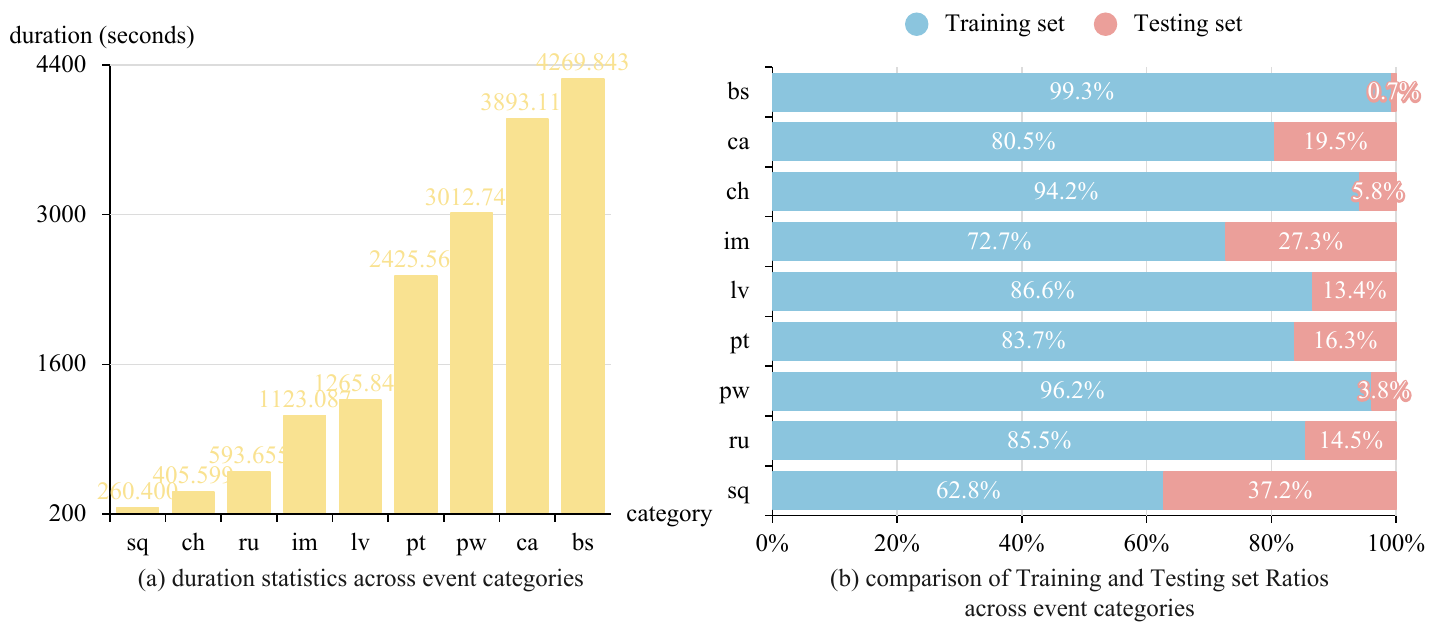}
\vspace{-10pt}
\caption{Overview of event durations and training-testing ratios across categories: (a) event duration statistics, (b) ratios between training and testing sets.}
\label{fig:fig3}
\vspace{-10pt}
\end{figure}

In this work, we construct an integrated dataset following the methodology described in \cite{zhang2024unified}. The dataset comprises four acoustic scenes (``home'', ``residential area'', ``city center'', and ``office'') and 25 distinct sound event categories. We select nine event classes as target events: `bird singing (bs)', `car (ca)', `children (ch)', `impact (im)', `large vehicle (lv)', `people talking (pt)', `people walking (pw)', `rustling (ru)', and `squeaking (sq)'. The remaining event categories are treated as detection-irrelevant events to simulate open-environment. All audio samples are sourced from the TUT 2016 dataset \cite{mesaros2016tut} and TUT 2017 dataset \cite{mesaros2017tut}, each containing annotations for both sound events and acoustic scenes. Detailed annotation information is available on this website \footnote{https://www.ksuke.net/dataset}. To evaluate the effectiveness of our proposed ensemble confidence calibration method in open environments, we utilize the ``home'' and ``residential area'' scenes from the TUT 2016 dataset as the training set (in-domain data), while the ``city center'' scene from the TUT 2017 dataset and the ``office'' scene from the TUT 2016 dataset serve as the test set (out-of-domain data). The training set contains a total audio duration of 194 minutes, while the test set comprises 70 minutes of audio. All audio samples are clipped or concatenated into 10-second clips. Figure~\ref{fig:fig3} presents the total duration statistics of each target event, along with the corresponding ratios between training and test sets.

We employ F-score \cite{fscore} as the evaluation metric, which can be categorized into two types based on evaluation granularity: event-based F-score focuses on the detection performance of complete event instances and their temporal boundaries, while segment-based F-score divides audio recordings into fixed-length segments for frame-level classification evaluation. In this work, we adopt the frame length as the basic unit for segment division. Furthermore, both evaluation approaches utilize macro-average and micro-average strategies. Therefore, the SED performance in open-environment is comprehensively assessed through four metrics: event-based macro-average F-score (Ema-F1), event-based micro-average F-score (Emi-F1), segment-based macro-average F-score (Sma-F1), and segment-based micro-average F-score (Smi-F1).

\begin{table*}[t!]
\centering
\caption{Performance comparison of different models for SED in open-environment}
\label{tab:performance_comparison}
\renewcommand{\arraystretch}{1.3}
\begin{tabular}{c|lcccc}
\toprule
ID & Model & Ema-F1 score (\%) $\uparrow$ & Emi-F1 score (\%) $\uparrow$ & Sma-F1 score (\%) $\uparrow$ & Smi-F1 score (\%) $\uparrow$ \\
\midrule
P1 & CRNN & 7.74 & 10.89 & 15.18 & 37.28 \\
P2 & EOW-Softmax & 9.07 & 13.00 & 17.91 & 43.34 \\
P3 & CRNN w/ avg. linear ensemble & 8.49 & 11.98 & 16.75 & 40.56 \\
P4 & EOW-Softmax w/ calibrated ensemble & \cellcolor[HTML]{ebebeb}\textbf{11.80} & \cellcolor[HTML]{ebebeb}\textbf{17.18} & \cellcolor[HTML]{ebebeb}\textbf{23.51} & \cellcolor[HTML]{ebebeb}\textbf{57.92} \\
\bottomrule
\end{tabular}
\end{table*}

\subsection{Implementation Details}

For data preprocessing, we use a sliding window to extract log mel-spectrogram features as audio representations. Each frame has a length of 128 milliseconds, and the window moves forward by 16 milliseconds. We apply soft mixup, a data augmentation method where two samples are mixed using a weight drawn from a Beta distribution with $\alpha = 0.2$ and $\beta = 0.2$. During training, we use the Adam optimizer with an initial learning rate of 0.001 and a batch size of 48. The training process runs for 200 epochs and starts with an exponential warmup during the first 50 epochs. We apply early stopping if the loss does not improve for 5 consecutive epochs. To ensure robustness, we train five models using different random seeds. The final model result is the average of the best checkpoints from these five runs, selected based on validation performance. For model ensembling, each candidate model is given an equal interpolation weight of 0.2. During inference, we apply median filtering with a window size of 7 to smooth predictions over time. Finally, we integrate the ensemble with EOW-Softmax for confidence calibration. The EOW-Softmax settings follow the same configuration as used in~\cite{wang2021energy}.

\section{Results and Discussion}

We present the results of four F-scores in Table~\ref{tab:performance_comparison}. All models were trained on acoustic scenes from ``home" and ``residential area" and tested on scenes from ``city center'' and ``office". Among the 25 event categories, nine were chosen as target events for detection. The other sixteen were used as background interference to simulate open-environment conditions.

We first compare the baseline system (P1) with an individual model trained using EOW-Softmax (P2). As shown in Table~\ref{tab:performance_comparison}, applying EOW-Softmax to the SOD branch leads to clear improvements in F-scores. This is due to the regularization effect of EOW-Softmax, which helps the model capture uncertainty in predictions and also reduces overfitting. Next, in method P3, we apply an ensemble strategy by averaging the outputs from five candidate models. This approach produces better results than the baseline (P1), but still falls short compared to the EOW-Softmax model (P2). The reason is that without confidence calibration, predictions with larger magnitudes dominate the average output. This imbalance can reduce the reliability of the final prediction, as discussed in prior work~\cite{kwok2025robust}. Finally, our proposed method (P4), which combines EOW-Softmax with model ensembling, achieves the best performance. It improves F-scores by 52.45\%, 57.76\%, 54.87\%, and 55.36\% over the baseline. These gains show that calibrating confidence allows the ensemble to better handle OOD inputs and improves robustness in open environments.

\begin{figure}[t!]
\centering
\includegraphics[width=0.85\columnwidth]{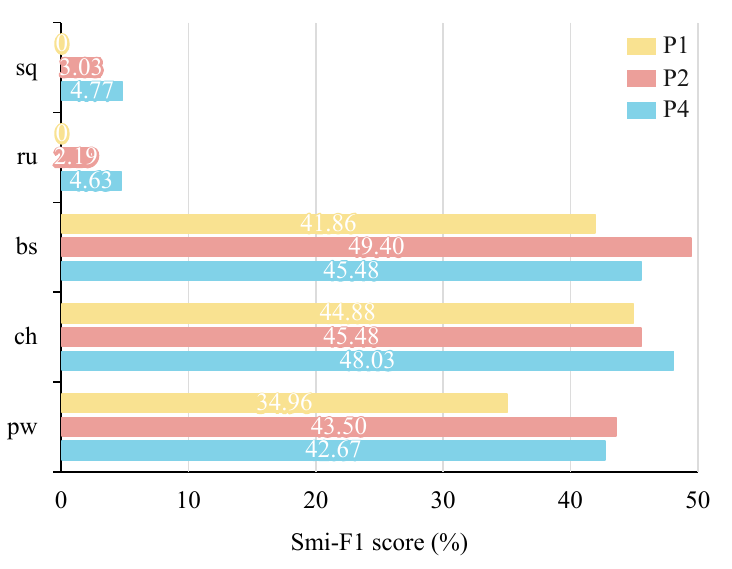}
\vspace{-10pt}
\caption{Smi-F1 score evaluation of different strategies for SED on challenging events, the strategy numbers correspond to the IDs listed in Table~\ref{tab:performance_comparison}.}
\label{fig:fig4}
\vspace{-5pt}
\end{figure}

To further validate the performance in open-environment settings, we selected several challenging sound events from Figure~\ref{fig:fig4} and compared the results of P1, P2, and P4 (as listed in Table~\ref{tab:performance_comparison}). These events fall into two groups: (1) events with imbalanced data, such as `bs', `ch', `pw', and (2) events with short durations and unclear features, such as `ru', `sq'. These cases reflect common difficulties in real-world scenarios, such as variation in event length, strong class imbalance, and high overlap in acoustic features. Figure~\ref{fig:fig4} shows the Smi-F1 scores for each method. Overall, models using EOW-Softmax outperform the baseline across all cases. Moreover, combining EOW-Softmax with ensembling (P4) gives the highest scores. This improvement comes from two main factors: first, EOW-Softmax makes training more stable by modeling uncertainty; second, calibrated ensembles more effectively combine predictions from multiple models, resulting in higher overall accuracy.

\section{Conclusions}

In this work, we innovatively propose the adoption of model ensembles to address SED in open-environment. Through experimental evaluation, we demonstrate that the introduced EOW-Softmax effectively regularizes the training process and performs confidence calibration during the model ensembles stage, thereby significantly improving the overall performance of SED. Future research can extend the application of this technique to more challenging OOD scenarios.

\clearpage

\printbibliography
\balance
\end{document}